# The case for the Universe to be a Quantum Black Hole

Antonio Alfonso-Faus

E.U.I.T. Aeronáutica, Plaza Cardenal Cisneros s/n, 28040 Madrid, SPAIN

E-mail: aalfonsofaus@yahoo.es

**Abstract.** – We present a necessary and sufficient condition for an object of any mass *m* to be a quantum black hole (q.b.h.): "*The product of the cosmological constant $\Lambda$ and the Planck´s constant $\hbar$, $\Lambda$ and $\hbar$ corresponding to the scale defined by this q.b.h., must be of order one in a certain universal system of units*". In this system the numerical values known for $\Lambda$ are of order one in cosmology and about $10^{122}$ for Planck´s scale. Proving that in this system the value of the cosmological $\hbar_c$ is of order one, while the value of $\hbar$ for the Planck´s scale is about $10^{-122}$, both scales satisfy the condition to be a q.b.h., i.e. $\Lambda\hbar \approx 1$. In this sense the Universe is a q.b.h..We suggest that these objects, being q.b.h.´s, give us the linkage between thermodynamics, quantum mechanics, electromagnetism and general relativity, at least for the scale of a closed Universe and for the Planck´s scale. A mathematical transformation may refer these scales as corresponding to infinity (our universe) and zero (Planck´s universe), in a "scale relativity" sense.

Pacs: 04.70.-s, 97.60.Lf, 98.80.Q

## 1. - Introduction.

In an effort to link general relativity (using the cosmological constant $\Lambda$ and the gravitational constant *G*) with quantum mechanics (using Planck´s constant $\hbar$ and a fundamental particle mass *m*) in 1967 Zel´dovich [1] arrived at the following expression

$$\Lambda = 8\pi \frac{G^2 m^6}{\hbar^4} \qquad (1)$$

In this expression *m* is the mass of a fundamental particle. We generalize this mass *m* in (1) by allowing it to have different values depending on the scale that we are considering. For the scale of the Universe ~ $10^{56}$ gr, for the Planck´s scale ~ $10^{-5}$ gr. The corresponding sizes are ~ $10^{28}$ cm. for the Universe and ~ $10^{-33}$ cm. for the Planck´s scale. We will check that the relation (1) is satisfied for the scale of the Universe as well as for the Planck´s scale. It is also satisfied for the "mixture": the cosmological constant $\Lambda_c$ and the mass $m$ of a fundamental particle, which is the original approach used by Zel´dovich [1].

We define a q.b.h. as an object of mass $m$ such that its gravitational radius, $Gm/c^2$, where $G$ is the gravitational constant and $c$ the speed of light, is of the order of its *generalized* Compton wavelength. We define this generalized wavelength as the ratio between the "equivalent" Planck´s constant $\hbar_m$, for the scale of the q.b.h. under consideration, and the momentum $mc$. A generalized cosmological constant $\Lambda_m$ can also be defined as the square of the inverse of the gravitational radius of the mass $m$. We have proved elsewhere, Alfonso-Faus [2] that the product $\Lambda_m \hbar_m$ is a dimensionless number of order one *for any q.b.h.* (by taking $G = c^3$ that also satisfies the choice of $G = c = 1$). Therefore this product is scale invariant. Taking $G$ and $c$ also as scale invariants, to be checked *a posteriori,* we see from (1) that Planck´s constant $\hbar$ varies as the square of the scale, as defined by the mass $m$. From (1) (with $\Lambda\hbar \approx 1$) we get

$$\hbar_m \approx (mc)^2 = (Gm/c^2)^2 \approx 1/\Lambda_m \qquad (2)$$

These relationships are valid for any q.b.h., using the choice of units with $G = c^3$ that preserves the Einstein field equations [3]. The usual choice of $G = c = 1$ is included in our choice $G = c^3$. But our choice is more general because it allows $G$ and $c$ to be time dependent, not necessarily constant, as long as the relation $G(t) = c(t)^3$ is satisfied. With this relationship we ensure that the Einstein-Hilbert action, that dictates the dynamics of the gravitational physical metric, has a constant coefficient in front of the integral defining the action. We have shown elsewhere [2] that Planck´s constant $\hbar$, which corresponds to the quantum scale, cannot be equal to 1 for this scale once this choice ($G = c = 1$) is made. Its value is $\hbar \approx 1/10^{122}$. We define the scale as the ratio between mass, length and time of the Universe and the mass, length and time of the component we are considering. For example, if we use as the component of reference Planck´s q.b.h. units, then the scale is the dimensionless number ~ $10^{61}$, which is the same for the three dimensionless ratios of mass, length and time, for the two scales. We see that the speed of light $c$, being the ratio of a length and a time, is scale invariant and so is $G$ because we have chosen $G = c^3$.

## 2.- The Weinberg relation

In 1972 Weinberg [4] pointed out a linkage between cosmology (using *G* and the cosmological Hubble´s constant *H*) and quantum mechanics (using Planck´s constant *ℏ* and a fundamental particle mass *m*) with the expression

$$m^3 = A\frac{\hbar^2 H}{Gc} \quad (3)$$

where *A* is a numerical constant of order one. Since *G* and *c* do not depend on scale, while *ℏ* depends on the square of the scale *m*, we see now from (3) that the scale of *m* goes like the scale of time *t* (or *1/H*), or as the scale of length (*ct*). This is the reason why we have a meaningful scale between the Universe and Planck´s world, the number ~ $10^{61}$, the same for the three dimensionless ratios of mass, length and time. This must be a very significant "coincidence" because the scale factor is a very large number, 61 orders of magnitude. It is very unlikely that this 3-coincidence of a very large dimensionless number be of a random nature. It fits better if we think of it as being a hint for some kind of relic "inflation" that converts the Planck´s scale into the Universal one. This property strengthens the fractal condition of the Universe, Alfonso-Faus [5]. Combining (2) and (3) we get for a q.b.h. of mass *m*

$$mH \approx 1 \quad \text{or} \quad m \approx t_m \quad (4)$$

Here we reinterpret the Hubble´s constant to be the inverse of the characteristic time or "age" of the corresponding q.b.h. This "age" $t_m$ is the time light takes to travel the size of the q.b.h.. The striking result in (4) is that mass and time run together (this is the Mass-Boom effect, Alfonso-Faus [6]). Planck´s mass (~ $10^{-5}$ gr.) is equivalent to Planck´s time (~$10^{-44}$ sec.) while the mass of the Universe ( ~$5.10^{56}$gr.) is equivalent to its age (~$5.10^{17}$sec.), and of course the ratio mass/time is the same for both scales, the Universe and Planck´s world. This ratio is the dimensionless numerical number ~ $10^{39}$.

## 3. - Scale invariant and scale dependent properties

*Scale invariants are G, c and the product Λℏ (as well as temperature T to be seen later). Linear dependent with scale are the mass m, length (ct) and time (as well as energy, Boltzmann constant and entropy) . Planck´s constant ℏ goes like the square of the scale.*

***The contradictory value for $\Lambda$, a factor difference of about $10^{122}$ for cosmology and the standard model of particles, is now solved: The contradiction is due to the two different scales one has to consider. <u>It is just the square of the ratio of the two scales.</u>***

If we eliminate the product $Gm^3$ between the expressions (1) and (3) we get

$$\Lambda \approx \left(\frac{H}{c}\right)^2 \approx \frac{1}{(ct)^2} \qquad (5)$$

which is a well known relation for the cosmological constant $\Lambda_c$. If we take as the unit of length $ct = 10^{28}$ cm then $\Lambda_c \approx 1$. Here $t$ is the age of the Universe and $ct$ its visible size. With the condition $\Lambda \hbar \approx 1$ we get from (2) and (5) the following relations for a q.b.h.

$$\hbar_m \approx (mc)^2 \approx (ct_m)^2 \approx (Gm/c^2)^2 \qquad (6)$$

We clarify that $t_m$ is the time light takes to travel the gravitational radius $Gm/c^2$. Using Newton's laws, Alfonso-Faus [7], arrived at a relation between an effective gravitational cross section area $\sigma_m$, for a gravitational mass $m$, and the cosmological Hubble parameter $H$:

$$\sigma_m \approx \frac{Gm}{c}\frac{1}{H} \qquad (7)$$

This gravitational area gives the effective cross section for gravitational interaction, and it is very close to the square of the geometrical size of the object under consideration, Alfonso-Faus [8]. If we eliminate the ratio $G/H$ between (3) and (7) we get the following result:

$$\sigma_m \approx \left(\frac{\hbar_m}{mc}\right)^2 \qquad (8)$$

This relationship is not restricted to black holes of any kind. It implies that the gravitational cross section area of a mass $m$, as we have defined it as the effective area for gravitational interaction, is of the order of the square of its generalized Compton wavelength, a quantum mechanical property associated with a generalized $\hbar_m$. The gravitational cross section area of any mass $m$, such that $Nm \approx M$ ($M$ being the mass of the Universe given by $N$ objects of mass $m$), is of the order of its geometrical area, Alfonso-Faus [8]. If the mass $m$ is a q.b.h. then obviously this area is also of the order of the square of its gravitational radius. Then our claim is that relation (8) can be considered to be a link between gravitation and quantum mechanics. By equating (7) to (8) we recover Weinberg´s relation (3) that links cosmology and gravitation with quantum mechanics too.

## 4. - The law of the geometric mean

Combining (7) and (8), and using $H \approx 1/t$ for the Hubble parameter, we get

$$\frac{\hbar_m}{mc} \approx \sqrt{\frac{Gm}{c^2} \cdot (ct)} \qquad (9)$$

Again this relationship is not restricted to black holes of any kind. The expression (9) is a connection between quantum mechanics and general relativity, including cosmology that may be stated as follows:

*The generalized Compton wavelength of any gravitational mass m is of the order of the geometric mean between its gravitational radius, $Gm/c^2$, and the visible size of the Universe ct.*

For example, the value for the Compton wavelength of fundamental particles is $\sim 10^{-12}$ cm. and its gravitational radius $\sim 10^{-52}$ cm. The relation (9) then gives for the size of the Universe ct the value $\sim 10^{28}$ cm. as it should. The time $t$ must be the characteristic time for light to travel the size of the object. For a q.b.h. (9) gives then the following results

$$L = \hbar_m/mc = Gm/c^2 = ct = mc = constant \qquad (10)$$

The constancy here comes from the constancy of momentum $mc$ and, of course, from the constancy of the angular momentum $\hbar_m$. From here we get

$$c = L/t \approx LH \qquad (11)$$

a result that has been presented elsewhere, Alfonso-Faus [8]. The Planck´s units of length, mass and time (a q.b.h.) are given by

$$(\hbar G/c^3)^{1/2} = 1.6 \times 10^{-33} \text{ cm}$$

$$(\hbar c/G)^{1/2} = 2.2 \times 10^{-5} \text{ g} \qquad (12)$$

$$(\hbar G/c^5)^{1/2} = 5.4 \times 10^{-44} \text{ sec}$$

It is well known that the simplest combination of the three constants of physics $c$ (the speed of light, relativity), $G$ (Newton´s constant, gravitation) and $\hbar$ (Planck´s constant, quantum mechanics) defines a particle with the physical properties given in (12). This particle has an energy $\hbar v$, similar to a photon of frequency $v$ with a value approximately equal to the inverse of the Planck time in (12). We can easily see that from (12) $Gm^2$ is just $\hbar c$ so that this *quantum black hole* may be considered to be a particle of mass $m$ or to be equivalent to a photon of energy equal to its self gravitational potential. This energy may be considered also to be of the order of the electric potential of a charge $\sim 10\ e$, the charge of the electron being $e$ or the scaled one. Since Planck´s constant runs like the square of the scale, the charges run like the scale (this is a generalization of the fine structure constant $\hbar c \sim e^2$. A quantum black hole connects gravitation (general relativity), electromagnetism and quantum mechanics. Being a photon it cannot radiate in the Hawking well known blackbody emission, certainly not for a closed Universe. Therefore they may be thought of as universes in themselves, being stable and probably charged, Alfonso-Faus [9]. On the other hand the Planck´s q.b.h. in (12) is immediately converted into the scale of the Universe by using $\hbar_c \approx 10^{122} \hbar$. Since $\hbar$ is $\approx 1/10^{122}$ we have $\hbar_c \approx 1$. Then from (6) $ct$ is $\approx 1$ and from (5) $\Lambda_c \approx 1$ too. Hence we finally get the relation $\hbar_c \Lambda_c \approx 1$ and the Universe is therefore a q.b.h. Its charge may be $\sim 10^{61} e$, that is able to balance its self gravitation, Alfonso-Faus [9].

We can now generalize the Planck´s quantum black hole given in (12) to any mass m. Since $G$ and $c$ cannot change with scale, only $\hbar$ can do it as shown in [2]. One has to think in terms of a quantum black hole as defined by any mass $m$ linked to its corresponding Planck´s constant $\hbar_m$ and through the scale given by its gravitational radius. For the whole Universe its mass M $\approx 10^{56}$ grams is linked to its corresponding Planck´s constant $\hbar_c \approx 10^{122} \hbar$, as found in [2], its scale being $ct \approx 10^{28}$ cm. Here the factor $10^{122}$ is just the square of the ratio of the two scales: the Universe and the Planck´s scale. This solves the well known $\Lambda$ problem. A mathematical transformation of scales can make a correspondence between these two limits: the mathematical constructs of infinity (our universe) and zero (Planck´s universe), Nottale [10].

## 5. - Time scale for oscillations

The characteristic oscillation time in a quantum black hole is given by the ratio of its size $Gm/c^2$ to the speed of light $c$, i.e.

$$1/v_g \approx Gm/c^3 \qquad (13)$$

We have seen that we are free to choose a system of units with $G/c^3 = 1$. The advantage of this choice is that, whatever the possible changes in time for $G$ or $c$, the derivation of the Einstein field equations from the action principle is ensured, thanks to the constancy of the factors in front of the action integrals, Alfonso-Faus [6]. We refer this choice of units as a universal system. Then from (8) we see that the characteristic time $\tau_g$ is given by

$$\tau_g \approx 1/v_g \approx m \qquad (14)$$

This is equivalent to relation (4). For the Universe the relation (14) implies $M = t$, the Mass-Boom effect presented elsewhere [7].

## 6.- Thermodynamics ($tT = constant$)

It is well known that quantum sizes of quantum particles of mass $m$ are of the order of their Compton wavelength: $\hbar/mc$. Also a characteristic time $\tau$ is obtained dividing this length by the speed of light, which is the same as the ratio of $\hbar$ to the relativistic energy $mc^2$. The value of $\hbar \approx 1/10^{122}$, found in [2], is valid for fundamental particles, like protons, as well as the Planck´s quantum black hole as given in (7). We have seen that for the scale of the Universe of mass $M$ the corresponding Planck´s constant is $\hbar_c \approx 10^{122} \hbar \approx 1$. We have made the choice $G = c^3$, which includes the usual choice of $G = c = 1$, but it is not possible to make also the widespread choice of $\hbar = 1$ for the quantum world. Such a choice is only valid for the whole universe. This hints that every scale has its corresponding equivalent Planck´s constant, $\hbar_m$. Then we can equate the three types of energies corresponding to relativity, quantum mechanics, electromagnetism and thermodynamics:

$$mc^2 \approx \hbar/\tau \approx \hbar v \approx k\, T \qquad (15)$$

where $k$ is a generalized Boltzmann constant. Then we have the relation

$$\hbar/k \approx \tau T \qquad (16)$$

Here we see that temperature is scale invariant but varies inversely proportional to time. It is interesting to note that the time parameter appears to have the same statistical nature as the thermodynamic temperature. and the same for the mass. For the entropy, as given by Hawking and Bekenstein for a black hole, we can use the relation (2) to arrive at

$$S \approx k/(\hbar_m c) \cdot Gm^2 \approx k \qquad (17)$$

The entropy of a q.b.h. is constant, time independent, and it is equivalent to the Boltzmann constant k for the scale under consideration. For our universe it is given by

$$S_u \approx k_u \approx 10^{61} k_p \qquad (18)$$

where $k_p$ is the usual Boltzmann constant. It represents the entropy of a Planck´s q.b.h. Then the extensive property of entropy is fulfilled. The entropy of the Universe in (18) can also be expressed as (using (16))

$$S_u \approx Mc^2/T \approx (Mc^2/\hbar_c) k_u \tau \approx k_u \qquad (19)$$

Then we have

$$\hbar_c \approx Mc^2 \tau \approx 1 \qquad (20)$$

This expression results from having the size of the Universe $c\tau = L = 1$. Hence we also have

$$Mc \approx 1 \qquad (21)$$

and since $M \approx \tau$ we finally get

$$c \approx 1/\tau \qquad (22)$$

that has been derived elsewhere, Alfonso-Faus [7]. The speed of light runs inversely proportional to the time scale.

## 7. - Conclusions

A necessary and sufficient condition for a mass *m* to be a q.b.h. is that the product $\Lambda\hbar$ be of order 1 in a certain system of units defined by the condition $G = c^3$. The constants $\Lambda$ and $\hbar$ have a value defined by the scale of the q.b.h. Each q.b.h. defines a scale for mass, length and time such that these properties have the same scale factor between them when comparing two q.b.h.´s. These objects are seen to be the junction of quantum mechanics, general relativity, electrodynamics and thermodynamics. They are universes in themselves as our own. No Hawking emission from them is possible because each q.b.h. can be interpreted to be equivalent to just one photon. Scale relativity can make the association of the mathematical infinity, versus our universe, and the mathematical zero, versus the Planck universe.

## 8. – References

[1] Zel'dovich, Ya.B., Pis'ma Zh. Eksp. Teor. Fiz., 1967, vol, **6**, p. 883.

[2] Alfonso-Faus, A., Astrophys. Space Sci., "Artificial contradiction between cosmology and particle physics: the Λ problem. (2009) 321:69-72 and [arXiv:0811.3933](arXiv:0811.3933) (v2) April (2009)

[3] Alfonso-Faus, A., "Zeldovich Λ and Weinberg Relation: An explanation for the Cosmological Coincidences", Astrophysics and Space Science, 318:117-123, (2008) and arXiv:0803.1309 (v5) November 2008.

[4] Weinberg, S., *Gravitation and Cosmology*, John Wiley and Sons, New York, 1972.

[5] Alfonso-Faus, A., "Fractal universe and the speed of light: Revision of the universal constants" arXiv: 0905.3966, May 2009, and New Advances in Physics, 2(2), Sept. 2008, pp 109-113.

[6] Alfonso-Faus, A., "Entropy in the Universe: A New Approach", Entropy, 2000, **2**, 168-171, [www.mdpi.org/entropy/](www.mdpi.org/entropy/), and "Structure of the Universe: Generalization of Weinberg´s relation" arXiv:0805.4762 (2008).

[7] Alfonso-Faus, A., [arXiv:0710.3039](arXiv:0710.3039) "The Speed of Light and the Hubble Parameter: The Mass-Boom Effect" Presented at the 9th Symposium on "Frontiers of Fundamental Physics", 7-9 Jan. 2008, University of Udine, Italy. AIP Conf. Proc. 1018:86-93, 2008; and Astrophysics and Space Science, 315:25-29, 2008.

[8] Alfonso-Faus, A., [arXiv:0805.4762](arXiv:0805.4762) "Structure of the Universe: Generalization of Weinberg relation" May 2008.

[9] Alfonso-Faus, A., arXiv:0903.5037 "On the nature of the outward pressure in the Universe". March 29, 2009. Also to appear in the next issue of the New Advances in Physics Journal.

[10] Nottale, L., 1992, Int. J. Mod. Phys. A7, 4899-4936 "The theory of scale relativity".